\documentclass[amsmath,amssymb,aps,pre,reprint,showpacs,groupedaddress,10pt]{revtex4-1}
\pdfoutput=1
\usepackage{graphicx}
\usepackage{dcolumn}
\usepackage{bm}  
\usepackage[utf8]{inputenc}
\usepackage{textcomp}
\usepackage[T1]{fontenc}
\usepackage{lmodern}
\usepackage{mathrsfs}
\usepackage{hyperref}
\usepackage{color}

\begin{document}
\title[Local thermodynamics and the generalized Gibbs-Duhem equation]{Local thermodynamics and the generalized Gibbs-Duhem equation in systems with long-range interactions}

\author{Ivan Latella}
\email{ilatella@ffn.ub.edu}
\author{Agust\'in P\'erez-Madrid}
\affiliation{Departament de F\'{i}sica Fonamental, Facultat de F\'{i}sica, Universitat de Barcelona,
Mart\'{i} i Franqu\`{e}s 1, 08028 Barcelona, Spain}

\begin{abstract}
The local thermodynamics of a system with long-range interactions in $d$ dimensions is studied using the mean-field approximation.
Long-range interactions are introduced through pair interaction potentials that decay as a power law in the interparticle distance.
We compute the local entropy, Helmholtz free energy, and grand potential per particle in the microcanonical, canonical and grand canonical ensembles, respectively.
From the local entropy per particle we obtain the local equation of state of the system by using the condition of local thermodynamic equilibrium.
This local equation of state has the form of the ideal gas equation of state, but with the density depending on the potential characterizing long-range interactions.
By volume integration of the relation between the different thermodynamic potentials at the local level, we find the corresponding equation satisfied by the potentials at the global level.
It is shown that the potential energy enters as a thermodynamic variable that modifies the global thermodynamic potentials.
As a result, we find a generalized Gibbs-Duhem equation that relates the potential energy to the temperature, pressure, and chemical potential.
For the marginal case where the power of the decaying interaction potential is equal to the dimension of the space, the usual Gibbs-Duhem equation is recovered.
As examples of the application of this equation, we consider spatially uniform interaction potentials and the self-gravitating gas.
We also point out a close relationship with the thermodynamics of small systems. 
\end{abstract}


\newcommand{\vect}[1]{\bm{#1}}
\newcommand{\e}{\mathrm{e}}
\newcommand{\dif}{\mathrm{d}}
\newcommand{\iunit}{\mathrm{i}}

\maketitle

\section{Introduction}
\label{Introduction}

A great variety of systems in nature are dominated by long-range interactions.
Examples are self-gravitating systems \cite{Antonov:1962,Lynden-Bell:1968,Thirring:1970,Padmanabhan:1990}, two-dimensional vortices \cite{Chavanis:2002:c}, nuclear physics \cite{Chomaz:2002}, and also toy models such as the Hamiltonian mean-field model \cite{Dauxois:2002}.
Among other physical properties, the remarkable thermodynamic behavior of systems with long-range interactions makes them extremely attractive.
They are intrinsically nonadditive and may have negative heat capacity in the microcanonical ensemble leading to ensemble inequivalence  \cite{Ruffo:2002,Bouchet:2005,Campa:2009}.
For very instructive reviews on the subject with more examples and phenomena we refer the reader to \cite{Campa:2009,Bouchet:2010}.

The relaxation towards thermodynamic equilibrium in long-range systems proceeds in a different manner than that of systems with short-range interactions \cite{Benetti:2012}.
In short-range systems, internal collisions drive the system to a state characterized by a Maxwell-Boltzmann distribution function.
To the contrary, in systems with long-range interactions with a large number of particles the evolution is collisionless and the system may remain trapped in a nonequilibrium quasistationary state that is not described by a Maxwell-Boltzmann distribution \cite{Benetti:2012,Levin:2008,Chavanis:2006,Teles:2011}.
The time that the system remains in this quasistationary state depends on the number of particles and diverges if the number of particles is infinite.
However, for a large but finite number of particles in a very long time limit, the system will evolve to a state of thermodynamic equilibrium characterized by a Maxwell-Boltzmann distribution \cite{Teles:2010} if such a state exists.
These states of thermodynamic equilibrium given by the theory of ensembles in Boltzmann-Gibbs statistical mechanics are those we will focus on throughout this paper.

Interactions in this kind of systems are characterized by slowly decaying pair interaction potentials that couple the constituent parts of the system at large distances.
Formally, a potential that decays as $1/r^\nu$ is said to be long range if $\nu\leq d$, where $d$ is the dimension of the embedding space \cite{Campa:2009}.
Systems having such interaction potentials are sometimes called systems with strong long-range interactions \cite{Bouchet:2010}.
The paradigmatic case of Newtonian gravity ($\nu=1$) has served as a basis for developing methods for studying an important part of the phenomenology concerning the thermodynamics of systems with long-range interactions.
In this regard, the isothermal spheres model has been widely used to study self-gravitating systems in the mean-field (MF) limit \cite{Padmanabhan:1990,Chavanis:2002:a}.
Although in the MF approach correlations are ignored, this model offers a mathematical tool for a suitable treatment of self-interactions in the system and turns out to be very accurate for a large number of particles \cite{deVega:2002:a,deVega:2002:b}, except near the critical points where the system collapses.
It is also well known that self-gravitating systems possess equilibrium states with negative heat capacity, provided the system is isolated (microcanonical ensemble).
Equilibrium in that case is ensured in a certain range in the space of parameters because isothermal spheres correspond to local maxima of the entropy with an extremely large lifetime that scales like the exponential of the number of particles \cite{Chavanis:2005}.
In the microcanonical ensemble the system becomes unstable when the heat capacity passes from negative to positive, leading to the gravothermal catastrophe \cite{Lynden-Bell:1968,Lynden-Bell:1999}.
When one of these systems is put in contact with a heat bath (canonical ensemble), the range in the space of parameters where the heat capacity is negative is replaced by a phase transition \cite{Thirring:1970}.
In the canonical ensemble, isothermal spheres correspond to states of local minima of the free energy and the isothermal collapse sets in when the heat capacity passes from positive to negative \cite{Chavanis:2002:a}.
The self-gravitating gas has also been studied in the grand canonical ensemble with the MF approach and Monte Carlo simulations.
There, the instability sets in at a critical value of the parameter controlling the state of the system that is different from the critical values in the microcanonical and canonical ensembles \cite{Horwitz:1977,Horwitz:1978,deVega:2002:a,deVega:2002:b}.
This illustrates the fact that different ensemble representations are, in general, inequivalent and the thermodynamic behavior of the system strongly depends on the control parameter used to specify its thermodynamic state.
In addition, systems with an attractive interaction potential $1/r^\nu$ with $0<\nu<3$ in three dimensions were also considered in the microcanonical ensemble \cite{Ispolatov:2001:a,Ispolatov:2001:b} and also there a critical energy was found below which these systems undergo a gravitational-like phase transition. Phase transitions in simplified models such as the ring model have also been studied \cite{Nardini:2009,Rocha:2011}.

Moreover, de Vega and S\'anchez obtained \cite{deVega:2002:b} the equation of state of the self-gravitating gas and pointed out that it was customary to assume it without any derivation.
To obtain the equation of state these authors took into account the condition of hydrostatic equilibrium and found that the system locally behaves as an ideal gas.
The same behavior is obtained if the self-gravitating gas possesses two or more kinds of particles with different mass \cite{deVega:2002:c} and for a system with arbitrary long-range interactions in the MF limit \cite{Chavanis:2011}.

In this framework, first, our aim here is to analyze the local thermodynamics of systems with strong long-range interactions in $d$ dimensions by computing the local thermodynamic potentials per particle in the MF limit.
The local equation of state is obtained by computing the local entropy per particle and using the condition of local thermodynamic equilibrium instead of assuming hydrostatic equilibrium.
Although in this case the two equilibrium conditions lead to the same result, both approaches are conceptually different.
Second, it is shown that the potential energy enters as a thermodynamic variable that modifies the global thermodynamic potentials.
As a result, we find a generalized Gibbs-Duhem equation that relates the potential energy to the thermodynamic variables of the system.

The rest of the paper is organized as follows.
In Sec.~\ref{mean:field:potentials} we consider a system with strong long-range interactions in $d$ dimensions and compute the local entropy, Helmholtz free energy, and grand potential per particle in the microcanonical, canonical, and grand canonical ensembles, respectively.
In Sec.~\ref{Local:relations} we obtain local relations satisfied by the thermodynamic potentials as well as the local equation of state.
In Sec.~\ref{Global:magnitudes} the global equation of state is obtained for a $d$-dimensional system and, by integrating the local relations, an equation satisfied by the set of global thermodynamic potentials is found.
We also obtain the generalized Gibbs-Duhem equation and consider some examples of its application.
In connection with the latter, we point out a close relationship existing between our treatment and Hill's \cite{Hill:1963} thermodynamics of small systems. 
Finally, in Sec.~\ref{Discussion} a discussion of our results is presented. We use units such that $k_\mathrm{B}=1$.

\section{Mean field local thermodynamic potentials}
\label{mean:field:potentials}

In this section we compute the MF local entropy, Helmholtz free energy, and grand potential per particle in the microcanonical, canonical, and grand canonical ensembles, respectively.
Here we follow a standard approach in the formulation of statistical mechanics (see, e.g., \cite{Chavanis:2006:b} and references therein) and hence thermodynamic potentials obtained in the different ensembles are also standard. From these potentials, the formulation of the thermodynamics at the local level is performed in Sec.~\ref{Local:relations} and exploited in Sec.~\ref{Global:magnitudes}.

Since the long-range character of the interactions depends on the dimension of the embedding space, we consider that the system is $d$ dimensional.
Interactions are introduced through a long-range pair interaction potential $\phi({\vect{q}}_i,{\vect{q}}_j)$, which depends on ${\vect{q}}_i$ and ${\vect{q}}_j$, the positions of particles $i$ and $j$.
It is formally assumed that at large enough distances the potential behaves as
\begin{equation}
\phi_{ij}\equiv\phi({\vect{q}}_i,{\vect{q}}_j)=\kappa |\vect{q}_i-\vect{q}_j|^{-\nu} ,
\label{interaction:potential}
\end{equation}
where $\kappa$ is a coupling constant and $0\leq\nu\leq d$.
The potential must be regularized at short distances in order to avoid divergences and consistently define the statistical mechanics of the system \cite{Chavanis:2006:b}. Thus here we consider implicitly a small distance cutoff that will be taken to be zero once the MF equilibrium configurations are obtained in the large $N$ limit \cite{Horwitz:1977,deVega:2002:a,Chavanis:2006:b}. In this way, short-range interactions are completely negligible in the MF limit and quantities such as the potential energy exist and are finite. Throughout the paper the potential energy has to be understood as an unknown or unevaluated quantity and in Sec.~\ref{Global:magnitudes} explicit examples are considered.
The power law (\ref{interaction:potential}) for the interaction will also be utilized in Sec.~\ref{Global:magnitudes} to derive global thermodynamic properties of the system.

\subsection{Local entropy: microcanonical ensemble}
\label{microcanonical:ensemble}

Consider a system of $N$ classical pointlike particles of equal mass $m$ enclosed in a container of $d$-dimensional volume $V\sim L^d$, $L$ being a characteristic length defining the size of the system.
Positions and momenta of particles are respectively described by
$\vect{q}_i=(q_i^1,q_i^2,\dots,q_i^d)$
and
$\vect{p}_i=(p_i^1,p_i^2,\dots,p_i^d)$, $1\leq i\leq N$.
The Hamiltonian of the system is
$H_N(\vect{q}_i,\vect{p}_i)=E_0+W$,
where the kinetic and potential energies $E_0$ and $W$, respectively, are given by
\begin{equation}
E_0=\sum_i^N\frac{|\vect{p}_i|^2}{2m}\qquad\mbox{and}\qquad W=\sum_{i>j}^N \phi_{ij}.
\end{equation}

In the microcanonical description the state of the system is characterized by a fixed value of the energy $E$ and the number of accessible microstates is given by
$\Sigma(E)=(2\pi\hbar)^{-dN}(N!)^{-1}\int_{E>H_N} \prod_{i=1}^N\dif^d\vect{p}_i\dif^d\vect{q}_i$.
The domain of spatial integrations extend over the $d$-dimensional volume $V$ and there are no restrictions on the domain of the momentum; this will always be the case throughout the paper unless another domain is specified.
Introducing $\varepsilon\equiv2\pi\hbar^2m^{-1}V^{-2/d}$ and integrating over momentum leads to
\begin{equation}
\begin{split}
\Sigma(E)&=\frac{1}{N!\Gamma\left(\frac{dN}{2}+1\right)} \int \frac{\dif^{dN}\vect{q}}{V^N}\\
&\qquad \times\,\left(\frac{E}{\varepsilon}	-\sum_{i>j}^N \frac{\phi_{ij}}{\varepsilon}\right)^{dN/2}
\theta\left(\frac{E}{\varepsilon}-\sum_{i>j}^N \frac{\phi_{ij}}{\varepsilon}\right) ,
\label{number:microstates}
\end{split}
\end{equation}
where $\dif^{dN}\vect{q}\equiv \prod_{i=1}^N\dif^d\vect{q}_i$, $\theta(x)$ is the Heaviside step function and $\Gamma(x)$ is the Gamma function.
Using the integral representation \cite{deVega:2002:a}
\begin{equation}
x^\gamma\theta(x)=\frac{\Gamma(\gamma+1)}{2\pi}\int_{-\infty}^{\infty}\dif k\ \frac{\exp\left(\iunit k x\right)}{(\iunit k)^{\gamma+1}} ,
\end{equation}
with $\iunit k=\varepsilon\beta$ and defining
\begin{equation}
\e^{J(\beta)}\equiv \frac{1}{N!}\int \frac{\dif^{dN}\vect{q}}{V^N}\exp\left(-\beta\sum_{i>j}^N \phi_{ij}-\frac{dN}{2}\ln\left(\varepsilon\beta\right)\right),
\label{integral}
\end{equation}
the number of microstates becomes
\begin{equation}
\Sigma(E)= \varepsilon\int_{-\iunit\infty}^{\iunit\infty}\frac{\dif\beta}{2\pi\iunit} \exp\left[\beta E+J(\beta) -\ln\left(\varepsilon\beta\right)\right] . 
\label{number:microstates:3}
\end{equation}

To evaluate (\ref{integral}), we assume $N\gg1$ and use the MF approach following Ref.~\cite{deVega:2002:a}. In this way, (\ref{integral}) can be written as a functional integral over the number density $n(\vect{x})$, $\vect{x}$ being a point in one-particle configuration space. We have (see the Appendix)
\begin{equation}
\e^{J(\beta)}=\int_{-\iunit\infty}^{\iunit\infty}\frac{\dif\alpha}{2\pi\iunit}\int\mathcal{D}n\,\exp\left(\hat{J}[n;\alpha,\beta]\right) , 
\label{exp:J}
\end{equation}
where we have introduced
\begin{equation}
\begin{split}
\hat{J}[n;\alpha,\beta]\equiv &\alpha \left(N-\int n(\vect{x})\, \dif^d\vect{x}\right)\\
&-\int n(\vect{x})\ln\frac{n(\vect{x})V}{\e}\, \dif^d\vect{x}\\
&- \frac{1}{2}\beta\int n(\vect{x})n(\vect{x}')\phi(\vect{x},\vect{x}')\,\dif^d\vect{x}\,\dif^d\vect{x}'\\
&-\frac{d}{2}\ln\left(\varepsilon\beta\right)\int n(\vect{x})\, \dif^d\vect{x} .
\label{functional:J}
\end{split}
\end{equation}
Since the microcanonical entropy is given by $S(E)=\ln \Sigma(E)$, substituting (\ref{exp:J}) in (\ref{number:microstates:3}) yields
\begin{equation}
\e^{S(E)}= \varepsilon\int_{-\iunit\infty}^{\iunit\infty}\frac{\dif\alpha}{2\pi\iunit}\int_{-\iunit\infty}^{\iunit\infty}\frac{\dif\beta}{2\pi\iunit}
\int \mathcal{D}n\, \exp\left(\hat{S}[n;\alpha,\beta]\right) ,
\label{exp:entropy}
\end{equation}
where
\begin{equation}
\hat{S}[n;\alpha,\beta]\equiv \beta E + \hat{J}[n;\alpha,\beta] .
\label{exponent}
\end{equation}
In the definition of $\hat{S}$ we have neglected the term
$\ln\left(\varepsilon\beta\right)$
coming from the exponential in (\ref{number:microstates:3}), since it does not contribute in the large $N$ limit.
This is the same as assuming that the microcanonical entropy is given by the logarithm of the number of microstates or, equivalently, by the logarithm of the density of states.
To proceed further, (\ref{exp:entropy}) will be evaluated using the saddle point approximation.
The value of the integral will be then given by the exponential of
$\hat{S}[n_\mathrm{s};\alpha_\mathrm{s},\beta_\mathrm{s}]$,
where
$n_\mathrm{s}(\vect{x})$, $\alpha_\mathrm{s}$, and $\beta_\mathrm{s}$
are the number density and the value of the parameters that maximize
$\hat{S}[n;\alpha,\beta]$.
Accordingly, the microcanonical entropy can be approximated by
$S(E)\approx\hat{S}[n_\mathrm{s};\alpha_\mathrm{s},\beta_\mathrm{s}]$.
In order to simplify the notation, in what follows we will omit the subscript $\mathrm{s}$ since we will only consider the stationary solutions given by the saddle point approximation.

The parameters $\alpha$ and $\beta$ in (\ref{exp:entropy}) can be viewed as Lagrange multipliers that respectively restrict the value of the number of particles and the energy.
Indeed, the MF equations in the saddle point approximation given by the extremal condition
$\left(\delta_n;\partial_\alpha,\partial_\beta\right)\hat{S}[n;\alpha,\beta]=(0;0,0)$
yield
\begin{align}
N&=\int n(\vect{x})\; \dif^d\vect{x} ,
\label{number:particles:constrain}\\
 E&=\frac{d}{2\beta}\int n(\vect{x})\;\dif^d\vect{x}+W[n],
\label{energy:constrain}
\end{align}
with
\begin{equation}
W[n]= \frac{1}{2}\int n(\vect{x})n(\vect{x}')\phi(\vect{x},\vect{x}')\,\dif^d\vect{x}\,\dif^d\vect{x}',
\label{potential:energy}
\end{equation}
after computing the derivatives of $\hat{S}[n;\alpha,\beta]$ with respect to $\alpha$ and $\beta$.
Moreover, taking into account (\ref{functional:J}), variations of the number density in (\ref{exponent}) lead to
$\delta\hat{S}[n]=\delta\hat{J}[n]$
with 
\begin{equation}
\delta \hat{J}[n]=\int \left\{ -\beta\Phi(\vect{x})-\alpha-\ln\left[\frac{n(\vect{x})V}{(\varepsilon\beta)^{-d/2}}\right]\right\}\delta n(\vect{x})\, \dif^d\vect{x}, 
\label{variations:S}
\end{equation}
where the term containing the potential $\Phi(\vect{x})$ comes from the variation
$\delta W[n]=\int\Phi(\vect{x})\delta n(\vect{x})\, \dif^d\vect{x}$
and is given by
\begin{equation}
\Phi(\vect{x})=\int n(\vect{x}')\phi(\vect{x},\vect{x}')\,\dif^d\vect{x}'=\int \frac{\kappa n(\vect{x}')}{|\vect{x}-\vect{x}'|^{\nu}}\,\dif^d\vect{x}' .
\label{potential}
\end{equation}
In terms of $\Phi(\vect{x})$, the total potential energy takes the form
\begin{equation}
W[n]=\frac{1}{2}\int n(\vect{x})\Phi(\vect{x})\,\dif^d\vect{x} .
\label{potential:energy:2}
\end{equation}
Hence, by setting the functional derivative with respect to $n(\vect{x})$ to zero,
$\delta_n\hat{S}[n]=\delta_n\hat{J}[n]=0$,
it follows the coupling between the number density and the potential through the relation
\begin{equation}
n(\vect{x})=\lambda_T^{-d}\exp\left[-\beta\left(\Phi(\vect{x})-\mu\right)\right] ,
\label{number:density:micro}
\end{equation}
where we have introduced
$\lambda_T=V^{1/d}(\varepsilon\beta)^{1/2}$ and $\mu=-\alpha/\beta$.
We emphasize that the density strongly depends on the form of the interaction potential, i.e, on $\nu$, as can be seen from (\ref{potential}).
As a result, by substituting (\ref{number:particles:constrain}) and (\ref{energy:constrain}) in (\ref{functional:J}) and (\ref{exponent}), the extrema of $\hat{S}$ and consequently the microcanonical MF entropy become
\begin{equation}
S= \int n(\vect{x})\left[-\ln\left(n(\vect{x})\lambda_T^d\right)+\frac{2+d}{2}\right]\, \dif^d\vect{x} ,
\label{microcanonical:S}
\end{equation}
which is an integral over the volume and therefore the integrand can be interpreted as the density of entropy.
From this density of entropy one can construct the local entropy per particle and in this way, as we shall see below, information about the local thermodynamic nature of the system can be obtained.
Taking into account (\ref{microcanonical:S}), the local entropy per particle is given by
\begin{equation}
s(\vect{x})=-\ln\left(n(\vect{x})\lambda_T^d\right)+\frac{2+d}{2}
\label{local:entropy1}
\end{equation}
in such a way that the total MF entropy can be written as
\begin{equation}
S=  \int n(\vect{x}) s(\vect{x})\, \dif^d\vect{x} .
\end{equation}

In addition, using (\ref{functional:J}) and (\ref{exponent}), it can be checked that
$\beta=\left(\partial_E S\right)_{N,V}=1/T$, $T$ being the temperature, and that
$\mu=-T\left(\partial_N S\right)_{E,V}=-\alpha T$ is the chemical potential.
Thus $\lambda_T=\sqrt{2\pi\hbar^2\beta/m}$ is the thermal wavelength.

The microcanonical MF entropy (\ref{microcanonical:S}) is a solution of the saddle point equations and therefore an extremum of Eq.~(\ref{exponent}). 
In order to guarantee thermodynamic equilibrium, we will assume that there exists a certain range of parameters where $\hat{S}$ is a maximum and thus our analysis here must be understood to be restricted to such a range.
It is not difficult to see that such a range in the space of parameters does exist: If the temperature is high enough so that $\beta\Phi(\vect{x})\ll1$, the number density becomes $n\approx N/V$ and the stable global ideal gas behavior is recovered.

The critical values of the parameters that set the instability will depend on the interactions under consideration.
For instance, consider the case of $\nu=1$ and $\kappa=-G_\mathrm{N} m^2$ in $d=3$ with spherical symmetry, which corresponds to self-gravitating isothermal spheres (see Sec.~\ref{self-gravity}).
Here $G_\mathrm{N}$ is Newton's constant. Consider also that the system is placed in a box of radius $R$.
The density contrast defined by $\mathcal{R}\equiv n(0)/n(R)$ is usually taken as the parameter describing the thermodynamic state of the system and the critical point at which the MF breaks down in the microcanonical ensemble is $\mathcal{R}=709$ \cite{Antonov:1962,Lynden-Bell:1968,Horwitz:1977,Horwitz:1978,Padmanabhan:1990}.
Beyond this point there are no stable configurations and the system undergoes gravitational collapse, known as the gravothermal catastrophe \cite{Lynden-Bell:1968}.
Furthermore, by solving the isothermal sphere equation, or Emden equation, (\ref{Emden}), one sees that $-ER/(N^2G_\mathrm{N}m^2)$ never exceeds $0.335$ for any equilibrium gas sphere \cite{Lynden-Bell:1968,Lynden-Bell:1999}, thus leading to a critical radius $R_\mathrm{A}=-0.335N^2G_\mathrm{N}m^2/E$.
When $R=R_\mathrm{A}$, the density contrast is $709$ and no stable thermal equilibrium exists for $R>R_\mathrm{A}$.

\subsection{Local Helmholtz free energy: canonical ensemble}
\label{canonical:ensemble}

Here we will consider that the control parameter used to specify the equilibrium configurations is the inverse temperature $\beta$.
At fixed temperature one must use the canonical description and in this section we will compute the canonical partition function $Z(\beta)$ in the saddle point approximation.
In doing so, the canonical Helmholtz free energy $A(\beta)$ can be obtained and written as an integral over the volume, which allows us to identify the local free energy.

The canonical partition function can be obtained from the microcanonical density of states
$\omega_E=\partial_E\Sigma(E)$
by computing its Laplace transform.
The density of states is then given by the inverse Laplace transform of the partition function
\begin{equation}
\omega_E= \int_{-\iunit\infty}^{\iunit\infty}\frac{\dif\beta}{2\pi\iunit} Z(\beta)\,\e^{\beta E} . 
\label{omega1}
\end{equation}
From (\ref{number:microstates:3}) one gets
\begin{equation}
\frac{\partial}{\partial E}\Sigma(E)=\int_{-\iunit\infty}^{\iunit\infty}\frac{\dif\beta}{2\pi\iunit}
\e^{J(\beta)}\,\e^{\beta E} .
\label{omega2} 
\end{equation}
Therefore, the comparison of both expressions (\ref{omega1}) and (\ref{omega2}) leads to the identification
$Z(\beta)=\e^{J(\beta)}$
and thus one sees that the function $J(\beta)$ defined in (\ref{integral}) is the Massieu function related to the Helmholtz free energy via $J(\beta)=-\beta A(\beta)$.
Since the canonical partition function is given by (\ref{exp:J}), the major part of the work has already been done in writing down the microcanonical description.
It is clear that the extremal condition $\left(\delta_n;\partial_\alpha\right)\hat{J}[n;\alpha,\beta]=(0;0)$ allows one to get an approximate expression for the Massieu function as $J(\beta)\approx\hat{J}[n;\alpha,\beta]$, where, as before, $n(\vect{x})$ and  $\alpha$ are given by the saddle point equations.
The saddle point equation associated with the parameter $\alpha$ constrains the number of particles leading to the condition (\ref{number:particles:constrain}).
Of course, in the canonical description only the temperature is fixed and there is no second Lagrange multiplier enforcing the condition~(\ref{energy:constrain}).
Instead, according to the saddle point approximation and using (\ref{functional:J}), one sees that the mean value of the Hamiltonian
$\langle E\rangle=-\partial_\beta \ln Z\approx-\partial_\beta \hat{J}\equiv \bar{E}$
becomes
\begin{equation}
\bar{E}=\frac{d}{2\beta}\int n(\vect{x})\;\dif^d\vect{x}+W[n].
\label{canonical:energy}
\end{equation}

In order to obtain the relation between the number density and the interaction potential, variations with respect to the number density of $\hat{J}[n]$ have to be computed.
This task has also been done in the previous section and $\delta \hat{J}[n]$ is given by expression (\ref{variations:S}).
As a result, the distribution of particles that maximizes the microcanonical entropy also maximizes the Massieu function in the canonical ensemble and the number density is then given by (\ref{number:density:micro}).
Hence, using (\ref{functional:J}), (\ref{number:particles:constrain}) and (\ref{potential:energy:2}), the saddle point approximation allows one to write the canonical MF Helmholtz free energy $A=-T J$ as an integral over the volume:
\begin{equation}
A= \int n(\vect{x})\left\{T\left[\ln\left(n(\vect{x})\lambda_T^{d}\right)-1 \right]+\frac{1}{2}\Phi(\vect{x})\right\}\,\dif^d\vect{x} .
\label{canonical:A}
\end{equation}
Therefore, the MF local Helmholtz free energy is readily identified,
\begin{equation}
a(\vect{x}) =T\left[\ln\left(n(\vect{x})\lambda_T^{d}\right)-1 \right]+\frac{1}{2}\Phi(\vect{x}) ,
\label{canonical:local:A}
\end{equation}
since
\begin{equation}
A= \int n(\vect{x})a(\vect{x})\,\dif^d\vect{x} . 
\end{equation}

The canonical MF entropy can be obtained via a Legendre transformation of the canonical MF free energy
$S=\beta(\bar{E}-A)$
and, as a consequence of the saddle point approximation \cite{Chomaz:2002}, it coincides with the microcanonical MF entropy (\ref{microcanonical:S}) only when both ensembles lead to the same description of the state of the system.
To the contrary, Legendre transformations cannot be applied to relate thermodynamic potentials from different ensembles if the ensembles are not equivalent. It must be stressed that the inequivalence of ensembles arises because the thermodynamic potentials become convex in a certain region of parameter space. If the entropy-energy curve presents a convex intruder, the microcanonical and the canonical ensembles are not equivalent \cite{Campa:2009}.

The MF solution in the canonical ensemble is valid only in the range of parameters where $\hat{J}$ is maximum.
Turning back to the example given in the previous section, self-gravitating isothermal spheres are well described with the MF in the canonical ensemble, provided $\mathcal{R}<32.1$ \cite{Horwitz:1977,Horwitz:1978,Chavanis:2002:a}.
For values of the density contrast larger than $32.1$, $\hat{J}$ is not a maximum and hence the MF description in the canonical ensemble breaks down at the critical value $\mathcal{R}=32.1$. At this critical point the system undergoes a gravitational collapse and this collapse cannot be described with the MF approach. What happens is that in the range in parameter space such that $32.1<\mathcal{R}<709$, the heat capacity is negative in the microcanonical ensemble. Configurations with negative heat capacity may be stable in the microcanonical ensemble, but are never realized in the canonical ensemble leading to a collapse or gravitational phase transition. Thus, for $32.1<\mathcal{R}<709$, the microcanonical and the canonical ensembles are not equivalent.

\subsection{Local grand potential: grand canonical ensemble}
\label{grand:canonical:ensemble}

Consider the canonical partition function as a function of the number of particles and its Laplace transform:
\begin{equation}
\mathcal{L}[Z](\alpha)=\int_0^\infty \dif N\, Z(N)\, \e^{-\alpha N} .
\label{laplace:Z}
\end{equation}
Taking $N$ as a continuous variable, the canonical partition function is constant with value $Z(N)$ for all $N$ in the interval $[N,N+1)$.
Therefore, the integral in (\ref{laplace:Z}) can be written as a sum of integrals over the intervals $[N,N+1]$ and subsequently integrated:
\begin{align}
\mathcal{L}[Z](\alpha)=&\sum_{N=0}^\infty Z(N)\int_N^{N+1} \dif \tilde{N}\, \e^{-\alpha \tilde{N}}\\
=&\frac{1-\e^{-\alpha}}{\alpha}\sum_{N=0}^\infty \e^{-\alpha N}Z(N) .
\label{laplace:Z:2}
\end{align}
From here, since $\alpha=-\beta\mu$, one recognizes the grand canonical partition function
\begin{equation}
\mathcal{Z}(\alpha)=\sum_{N=0}^\infty \e^{-\alpha N}Z(N) , 
\end{equation}
which we will take as the starting point to derive the thermodynamics of the system.
Introducing
$\zeta(\alpha)=[1-\exp(-\alpha)]/\alpha$
and applying the inverse Laplace transform to (\ref{laplace:Z:2}) one gets
\begin{equation}
Z(N)=\mathcal{L}^{-1}[\zeta\mathcal{Z}](N)= \int_{-\iunit\infty}^{\iunit\infty}\frac{\dif\alpha}{2\pi\iunit}\, \zeta(\alpha)\mathcal{Z}(\alpha)\e^{\alpha N} .
\end{equation}
Since in the previous section we obtained
\begin{equation}
Z(N)=\int_{-\iunit\infty}^{\iunit\infty}\frac{\dif\alpha}{2\pi\iunit}\int\mathcal{D}n\,\exp\left(\hat{J}[n;\alpha,\beta]\right) , 
\end{equation}
the grand canonical partition function reads
\begin{equation}
\mathcal{Z}=\int\mathcal{D}n\,\exp\left(\hat{J}[n;\alpha,\beta]-\alpha N-\ln \zeta(\alpha)\right) . 
\label{grand:canonical:partition:function}
\end{equation}
The term $\ln \zeta(\alpha)$ in the exponential above, which is a correction due to the discreteness of $N$, can be safely neglected in the large $N$ limit.
As before, we shall evaluate the integral (\ref{grand:canonical:partition:function}) using the saddle point approximation and hence
$\mathcal{Z}\approx\exp(\hat{J}-\alpha N)$.
In this case, only the number density such that
$\delta_n (\hat{J}[n]-\alpha N)=\delta_n \hat{J}[n]=0$
is required.
Thus one more time the number density is given by (\ref{number:density:micro}).
Furthermore, the mean value of the number of particles can be written as
$\langle N\rangle=-\partial_\alpha\ln\mathcal{Z}\approx-\partial_\alpha(\hat{J}-\alpha N )\equiv\bar{N}$
and using (\ref{functional:J}) one gets
\begin{equation}
\bar{N}= \int n(\vect{x})\;\dif^d\vect{x} .
\end{equation}
Analogously, the mean value of the energy takes the form
$\langle E\rangle =-\partial_\beta\ln\mathcal{Z}\approx -\partial_\beta(\hat{J}-\alpha N)=\bar{E}$,
where, as in the canonical case, $\bar{E}$ is given by (\ref{canonical:energy}).
Introducing the grand potential
\begin{equation}
\varOmega=-T\ln\mathcal{Z} , 
\end{equation}
the saddle point approximation yields
$-\beta\varOmega\approx\hat{J}-\alpha N$,
which by using (\ref{functional:J}) and (\ref{potential:energy:2}) leads to
\begin{equation}
\varOmega= \int n(\vect{x})\left\{T\left[\ln\left(n(\vect{x})\lambda_T^{d}\right)-1 \right]-\mu+\frac{1}{2}\Phi(\vect{x})\right\}\, \dif^d\vect{x} .
\label{grand:potential}
\end{equation}
Consequently, the MF local grand potential per particle reads
\begin{equation}
\omega(\vect{x})= T\left[\ln\left(n(\vect{x})\lambda_T^{d}\right)-1 \right]-\mu+\frac{1}{2}\Phi(\vect{x}) ,
\label{local:grand:potential}
\end{equation}
such that
\begin{equation}
\varOmega= \int n(\vect{x})\omega(\vect{x}) \, \dif^d\vect{x} .
\label{grand:potential2}
\end{equation}

The grand canonical MF Helmholtz free energy is obtained by a Legendre transformation of the form $A=\varOmega+\mu\bar{N}$, which turns out to be equal to the MF Helmholtz free energy in the canonical ensemble (\ref{canonical:A}) in the region of parameter space where both ensembles coincide.
With a subsequent Legendre transformation, we would obtain the same MF entropy as in the microcanonical ensemble
$S=\beta(\bar{E}-\mu\bar{N}-\varOmega)$ only when the grand canonical and the microcanonical ensembles are equivalent.
As mentioned before, this is a consequence of the saddle point approximation and these Legendre relations cease to be valid when the ensembles are not equivalent.
Although each ensemble has its own range of validity in the corresponding space of parameters, the thermodynamic potentials computed in the different ensembles have all the same form in the MF approach.
In the grand canonical ensemble the MF solution is valid only in the range of parameters such that $\hat{J}-\alpha N$ is maximum.
For self-gravitating isothermal spheres, the critical point where the MF description ceases to be valid in the grand canonical ensemble is $\mathcal{R}=1.58$ \cite{Horwitz:1977,Horwitz:1978}. At this critical point the system undergoes gravitational collapse in the grand canonical ensemble, which cannot be described by the MF. We see again that the stability of the system depends on the control parameters used to specify the state of the system. In addition, in this case the grand canonical MF entropy coincides with the microcanonical MF entropy only for $\mathcal{R}< 1.58$.

\section{Local relations and the local equation of state}
\label{Local:relations}

The formulation of thermodynamics in terms of local variables is well known (see, for instance, \cite{deGroot:1984,Glansdorff:1977}) and here we implement this formalism to systems with long-range interactions.
In the previous sections we wrote the thermodynamic potentials as integrals over the volume of the system, which leads to a natural definition of local quantities per particle.
Analogously, using (\ref{energy:constrain}) and (\ref{potential:energy:2}), the local energy per particle takes the form
\begin{equation}
e(\vect{x})=e_0+\frac{1}{2}\Phi(\vect{x}) ,  
\label{local:energy}
\end{equation}
where $e_0=\frac{d}{2}T$ is the local kinetic energy per particle, such that the total energy and kinetic energy are given by
$E=\int n(\vect{x}) e(\vect{x})\, \dif^d\vect{x}$ and $E_0=\int n(\vect{x}) e_0\, \dif^d\vect{x}$, respectively.
We also introduce the local volume per particle $v(\vect{x})=1/n(\vect{x})$, which obviously leads to $V= \int n(\vect{x}) v(\vect{x})\, \dif^d\vect{x}$.

Once local variables are defined, the local equation of state can be obtained.
In order to do that, it is useful to write the local entropy per particle in a more convenient way.
From (\ref{local:entropy1}) one obtains
\begin{equation}
 s(\vect{x})=\ln\left[v(\vect{x})\left(\frac{m}{d\pi \hbar^2}e_0\right)^{d/2}\right]+\frac{2+d}{2} ,
\end{equation}
which is a Sackur-Tetrode-like equation in $d$ dimensions formulated in terms of local variables. Long-range interactions are included in the local volume through its dependence on the interaction potential, thus leading also to an implicit dependence of the local entropy on the interaction potential.
The local entropy per particle is therefore explicitly obtained as a function of the local variables $e_0$ and $v$, $s=s(e_0,v)$.
Thus, in the equilibrium framework, one infers that the local internal energy only has contributions coming from kinetic degrees of freedom while self-interactions play the role of an external field that perturbs the gas.
In addition, local thermodynamic equilibrium implies the relations \cite{deGroot:1984,Glansdorff:1977}
\begin{equation}
\frac{1}{T}=\left(\frac{\partial s}{\partial e_0}\right)_v
\qquad\mbox{and}\qquad\frac{p}{T}=\left(\frac{\partial s}{\partial v}\right)_{e_0} , 
\label{local:equilibrium}
\end{equation}
where the second of these relations is the local equation of state of the system.
In our case, this local equation of state is indeed the one corresponding to an ideal gas,
\begin{equation}
p(\vect{x})=n(\vect{x})T ,
\label{local:equation:state}
\end{equation}
which is valid for any long-range pair interaction potential that can be suitably represented with the MF approach.
Since the MF local entropy per particle takes the same functional form in the microcanonical, canonical, and grand canonical ensembles, the above equation of state is valid in the three ensembles.
Note that the condition of local thermodynamic equilibrium together with expressions (\ref{local:equilibrium}) are formulated as a hypothesis in the framework of non-equilibrium thermodynamics, but here they are trivially satisfied provided the whole system is in equilibrium.
We emphasize that the local equation of state~(\ref{local:equation:state}) is usually derived by considering the condition of hydrostatic equilibrium \cite{deVega:2002:b,Chavanis:2011}; here it is derived on pure thermodynamic grounds from the local entropy.

To get more insight in the relation between local thermodynamic variables, we use (\ref{number:density:micro}) and write the chemical potential as
\begin{equation}
\mu=\mu_0(\vect{x})+\Phi(\vect{x}) ,
\label{chemical:potential}
\end{equation}
where $\mu_0(\vect{x})= T\ln\left(\lambda_T^dn(\vect{x})\right)$ possesses the same functional dependence on the number density as the chemical potential of the ideal gas, but with the density given by (\ref{number:density:micro}).
Also using (\ref{number:density:micro}), (\ref{chemical:potential}), and the local equation of state, one obtains
\begin{equation}
Ts(\vect{x})=[e_0+\Phi(\vect{x})]+p(\vect{x})v(\vect{x})-\mu .
\label{local:relation:2}
\end{equation}
The term in square brackets is the total energy of a particle at the point $\vect{x}$; locally the potential $\Phi(\vect{x})$ acts like an external field.
However, the difference between $\Phi(\vect{x})$ and an authentic external field becomes manifest when one sums the contribution of the whole system: Multiplying $[e_0+\Phi(\vect{x})]$ by the density and integrating over the volume does not give the total energy.
In order to obtain the total energy one must take into account that the total potential energy is due to self-interactions and this is the reason why one defines the local energy per particle according to (\ref{local:energy}) (the total potential energy is a functional quadratic in the density).
In this respect, using (\ref{local:energy}), Eq.~(\ref{local:relation:2}) can be alternatively written in the form
\begin{equation}
Ts(\vect{x})=e(\vect{x})+p(\vect{x})v(\vect{x})-\mu+\frac{1}{2}\Phi(\vect{x}) ,
\label{local:relation}
\end{equation}
where the last term highlights the fact that self-interactions are actually considered.
 
The local Helmholtz free energy per particle $a(\vect{x})$ can be obtained by the Legendre transformation
\begin{equation}
a(\vect{x})=e(\vect{x})-Ts(\vect{x})=\mu-p(\vect{x})v(\vect{x})-\frac{1}{2}\Phi(\vect{x}) .
\label{local:helmholtz}
\end{equation}
It can be checked that (\ref{local:helmholtz}) coincides with (\ref{canonical:local:A}).
The local Helmholtz free energy per particle can also be written in the form
$a(\vect{x})= a_0(\vect{x})+\frac{1}{2}\Phi(\vect{x})$, where $a_0(\vect{x})$ has the form of the local Helmholtz free energy per particle of an ideal gas at the point $\vect{x}$ and it is given by
\begin{equation}
a_0(\vect{x})= -T\left[\ln\left(v(\vect{x})\lambda_T^{-d}\right)+1 \right] .
\end{equation}
Hence it also satisfies
$a_0(\vect{x})=e_0-Ts(\vect{x})=\mu_0(\vect{x})-p(\vect{x})v(\vect{x})$.
We emphasize that $a_0(\vect{x})$ is a function of the local volume and hence depends on the interaction potential.
Moreover, the local grand potential (\ref{local:grand:potential}) is equivalent to the Legendre transformation
$\omega(\vect{x})=a(\vect{x})-\mu$
and with elementary manipulations it can be rewritten as
\begin{equation}
\omega(\vect{x})=-p(\vect{x})v(\vect{x})-\frac{1}{2}\Phi(\vect{x}) .
\label{local:grand:potential:2}
\end{equation}
The local grand potential per particle of an ideal gas takes the form
\begin{equation}
\omega_0=a_0(\vect{x})-\mu_0(\vect{x})=-p(\vect{x})v(\vect{x})=-T ,
\end{equation}
so it is related to $\omega(\vect{x})$ through
$\omega(\vect{x})=\omega_0-\frac{1}{2}\Phi(\vect{x})$.
By substituting (\ref{local:grand:potential:2}) in (\ref{grand:potential2}) and using the local equation of state one sees that the total grand potential is closely related to the kinetic and potential energies
\begin{equation}
\varOmega= -\int p(\vect{x}) \, \dif^d\vect{x}-W= -\frac{2}{d}E_0 -W .
\end{equation}

The discussion in Sec.~\ref{mean:field:potentials} about the correctness of Legendre transformations for global variables in connection with ensemble inequivalence is also valid here for local variables. Furthermore, it is worth noting that the Legendre transformation of a local variable with respect to the local volume, in general, does not agree with the corresponding Legendre transformation of the global variable with respect to the total volume.
The reason is that, in general, the pressure is not uniform.

\section{Global magnitudes and the generalized Gibbs-Duhem equation}
\label{Global:magnitudes}

Useful relations between global quantities can be obtained by integration of the local equations.
Multiplying both sides of (\ref{local:relation}) by $n(\vect{x})$ and integrating over the volume yields
\begin{equation}
TS=E+\frac{2}{d}E_0-\mu N +W .
\label{global:relation:1} 
\end{equation}
However, it would be desirable that the quantity $PV$ appears explicitly related to the other thermodynamic quantities.
In order to do that, the form of the pair interaction potential has to be taken into account and the global equation of state has to be considered. To derive this equation of state we follow Ref.~\cite{deVega:2002:a}.

We will consider that the container is a $d-1$ sphere of radius $R$ in such a way that particles are confined by the boundary condition
$\left(q_i^1\right)^2+\left(q_i^2\right)^2+\cdots+\left(q_i^d\right)^2\leq R^2$, $1\leq i\leq N$.
The zero sphere is the pair of endpoints of the line segment of length $2R$.
In addition, the volume of the system reads
$V=\pi^{d/2}R^d[\Gamma(d/2+1)]^{-1}$.

Thus, from (\ref{number:microstates}), using coordinates defined according to
$\vect{r}_i=\vect{q}_i/R$
and rescaling energies by introducing \cite{deVega:2002:a,Ispolatov:2001:b}
\begin{align}
\Lambda&\equiv \frac{ER^\nu}{|\kappa|N^2}=\frac{E}{|\kappa|N^2}\left(\frac{\Gamma(d/2+1)}{\pi^{d/2}}V\right)^{\nu/d} ,\\
\varphi_{ij}&\equiv \frac{R^\nu}{|\kappa|N^2} \phi(\vect{q}_i,\vect{q}_j)= \frac{1}{N^2}\frac{\kappa}{|\kappa|} |\vect{r}_i-\vect{r}_j|^{-\nu} ,
\end{align}
the entropy takes the form 
\begin{equation}
S=\ln \frac{V^{N(2-\nu)/2}}{\vartheta}+\ln \tilde{\Sigma}(\Lambda) . 
\end{equation}
Here
\begin{equation}
\vartheta=\left(\frac{2\pi\hbar^2}{N^2|\kappa|m}\right)^{dN/2} 
\left(\frac{\pi^{d/2}}{\Gamma(d/2+1)}\right)^{N(2-\nu)/2} 
\end{equation}
and
\begin{equation}
\begin{split}
\tilde{\Sigma}(\Lambda)=&\frac{1}{N!\Gamma(dN/2+1)}\int_D\dif^{dN}\vect{r}\\
&\times \left(\Lambda-\sum_{i>j}^N\varphi_{ij}\right)^{dN/2}\theta\left(\Lambda-\sum_{i>j}^N\varphi_{ij}\right) .
\label{sigma}
\end{split}
\end{equation}
Note that we assume that energy scales as $N^2$, as is well known in this kind of system \cite{Chavanis:2006:c}.
Note also that $\tilde{\Sigma}(\Lambda)$ depends on the volume only through $\Lambda$ since in the $\vect{r}_i$ coordinates,
which belong to a certain domain $D$, the limits of the integrals in (\ref{sigma}) are pure numbers.
Taking into account that
$(\partial_E \Lambda)_V=\Lambda/E$
and
$(\partial_V \Lambda)_E=\nu\Lambda/(d V)$,
one has
\begin{align}
\frac{1}{T}&=\left(\frac{\partial S}{\partial E}\right)_V =\frac{\Lambda}{E}\frac{\partial
\ln\tilde{\Sigma}(\Lambda)}{\partial \Lambda} ,\\
\frac{P}{T}&=\left(\frac{\partial S}{\partial V}\right)_E=\frac{N}{V}\left(1-\frac{\nu}{2}\right)+\frac{\nu \Lambda}{
dV}\frac{\partial \ln\tilde{\Sigma}(\Lambda)}{\partial \Lambda} ,
\end{align}
where
$P=p(\vect{x})|_{\vect{x}\; \in\partial V}$ is the pressure evaluated at the boundary of the system.
From the above equations and using that $E=E_0+W$ one gets
\begin{equation}
\frac{PV}{NT}= 1+\nu\frac{W}{dNT} ,
\label{equation:state}
\end{equation}
which is the exact microcanonical equation of state of the system.
However, in the MF limit this equation holds also in the canonical and grand canonical ensembles if mean values are taken such that $\bar{E}=E$ and $\bar{N}=N$.
Of course, the last statement is only valid in the domain of the space of parameters where the canonical and grand canonical ensembles are well defined and coincide with the microcanonical ensemble.
If the equivalence of ensembles does not hold, the corresponding equation of state has to be derived for each ensemble separately.
With this in mind, for simplicity, in what follows we will not distinguish between $E$ and $\bar{E}$ or $N$ and $\bar{N}$.

From the equation of state (\ref{equation:state}) the virial theorem is obtained, 
\begin{equation}
2E_0+\nu W=dPV ,
\label{virial} 
\end{equation}
which is particularly useful to express the relation between global quantities.
It is also useful to introduce the long-range parameter $\sigma$ defined as
\begin{equation}
\sigma\equiv \frac{d-\nu}{d} ,\qquad 0\leq\sigma\leq1 ,
\end{equation}
which together with (\ref{virial}) enables us to rewrite (\ref{global:relation:1}) as the Euler relation in the form
\begin{equation}
TS=E+PV-\mu N +\sigma W .
\label{global:relation:2} 
\end{equation}
The marginal case $\nu=d$ corresponds to systems with long-range parameter $\sigma=0$, so the above equation and the thermodynamic potentials obtained below reduce to the usual ones in short-range interactions thermodynamics.
The Helmholtz free energy, the grand potential, and the Gibbs free energy are readily obtained from (\ref{global:relation:2}) and take the form 
\begin{align}
A&=E-TS=\mu N-PV -\sigma W ,\\ 
\varOmega&=A-\mu N=-PV -\sigma W  ,\\
G&=A+PV=\mu N -\sigma W  .
\end{align}
The above expressions generalize previous results for the self-gravitating gas \cite{deVega:2002:a,deVega:2002:b}.
In addition, by differentiating (\ref{global:relation:2}) one gets
\begin{equation}
T\dif S=\dif E+P\dif V -\mu\dif N +\sigma\dif W -N\dif \mu-S\dif T+V\dif P
\label{differential_relation} 
\end{equation}
and since
$T\dif S = \dif E+P\dif V -\mu\dif N$
we must have
\begin{equation}
\sigma\dif W= S\dif T - V\dif P + N\dif \mu ,
\label{Gibbs-Duhem}
\end{equation}
which is the generalized Gibbs-Duhem equation satisfied by the long-range interacting systems we are considering here.
For systems with long-range parameter $\sigma=0$, i.e., the marginal case $\nu =d$, the usual Gibbs-Duhem equation is recovered.
From (\ref{Gibbs-Duhem}) we infer that when long-range interactions are present, $T$, $P$, and $\mu$ are independent.
Moreover, in view of (\ref{Gibbs-Duhem}), the following relations are obtained for $\sigma\neq0$:
\begin{subequations}
\begin{align}
S&=\sigma\left(\frac{\partial W}{\partial T}\right)_{P,\mu} ,\\
V&=-\sigma\left(\frac{\partial W}{\partial P}\right)_{T,\mu} ,\\
N&=\sigma\left(\frac{\partial W}{\partial \mu}\right)_{P,T} . 
\end{align}
\label{Gibbs-Duhem:relations}
\end{subequations}
The above expressions imply, for instance, that the entropy can be obtained from a derivative of the potential energy.
Since such thermodynamic relations are lacking for systems with long-range interactions, it would be interesting to check them by considering some solvable examples.
Next, we will consider the case where the interaction potential is spatially uniform and the case where the system is a self-gravitating gas.
After these two examples we will expose a common feature between systems with long-range interactions and small systems.

\subsection{Spatially uniform interaction potentials in $d$ dimensions}

These systems are characterized by $\nu =0$ such that the system is homogeneous and $\sigma=1$.
We simply have $\phi=\kappa$, hence $\Phi=k N$ and $W=\kappa N^2/2$.
In order to apply (\ref{Gibbs-Duhem:relations}), the potential energy must be written as a function of $T$, $P$, and $\mu$.
Using (\ref{number:density:micro}) and the equation of state (\ref{equation:state}), which in this case reads $PV=NT$, one obtains
$n=N/V=P/T=\lambda_T^{-d}\exp\left[-\beta(\kappa N-\mu)\right]$
and the number of particles becomes
\begin{equation}
N=\frac{1}{\kappa}\left[\mu-T\ln\left(\frac{P\lambda_T^d}{T}\right)\right] . 
\label{N:example:homogeneous}
\end{equation}
The potential energy then takes the form
\begin{equation}
W(T,P,\mu)=\frac{1}{2\kappa}\left[\mu-T\ln\left(\frac{P\lambda_T^d}{T}\right)\right]^2 .
\end{equation}
Taking into account that the local entropy per particle (\ref{local:entropy1}) is uniform in this case, whence $S=Ns$, it is straightforward to see that
\begin{align}
S&=\left(\frac{\partial W}{\partial T}\right)_{P,\mu}\nonumber\\
&=\frac{1}{\kappa}\left[\mu-T\ln\left(\frac{P\lambda_T^d}{T}\right)\right]\left[-\ln\left(\frac{P\lambda_T^d}{T}\right)+\frac{2+d}{2}\right] . 
\end{align}
The remaining two relations (\ref{Gibbs-Duhem:relations}) follow from $V=NT/P$ and (\ref{N:example:homogeneous}).

\subsection{Self-gravitating isothermal spheres in $d=3$}
\label{self-gravity}

The calculation of global thermodynamic quantities of self-gravitating isothermal spheres is well understood and here we will just consider it as an example to illustrate the relations (\ref{Gibbs-Duhem:relations}).
In particular, the first of these relations will be used to obtain the entropy of the self-gravitating gas.
We will write down only the necessary expressions to carry out our task and refer the reader to \cite{Padmanabhan:1990,Chavanis:2002:a,Chavanis:2005} for details.

In the self-gravitating gas we have $\nu=1$, thus $\sigma=2/3$, and $\kappa=-G_\mathrm{N} m^2$, where $G_\mathrm{N}$ is Newton's constant.
Also, Newtonian systems satisfy the Poisson-Boltzmann equation which, after introducing the dimensionless variables
$\xi=[4\pi|\kappa|\beta\, n(0)]^{1/2}x$ and $\psi=\beta\left(\Phi(x)-\Phi(0)\right)$,
becomes the Emden equation
\begin{equation}
\frac{1}{\xi^2}\frac{\dif}{\dif\xi}\left(\xi^2\frac{\dif}{\dif\xi}\psi\right)=\e^{-\psi} .
\label{Emden}
\end{equation}
Here $x=|\vect{x}|$ and $\beta=1/T$ while $n(0)$ and $\Phi(0)$ are the density and the potential at the origin.
Global thermodynamic magnitudes are expressed in terms of these functions evaluated at the boundary of the system, thus, for convenience, one introduces
$\xi_0=(4\pi |\kappa|\beta\, n(0))^{1/2} R$, $\psi_0=\psi(\xi_0)$ and $\psi'_0=\psi'(\xi_0)$,
where primes denote the derivative with respect to $\xi$ and $R$ is the radius of spherical container.
Moreover, with a suitable change of variables the Emden equation can be transformed into a first-order differential equation \cite{Chandrasekhar:1939}.
When such variables, usually denoted by $(v,u)$, are evaluated at $\xi=\xi_0$ and $\psi=\psi_0$, with $\psi'=\psi'_0$, they read
\begin{equation}
v_0=\xi_0\psi_0' \qquad\mbox{and}\qquad u_0=\frac{\xi_0\e^{-\psi_0}}{\psi_0'}
\end{equation}
and therefore satisfy
\begin{equation}
\frac{\dif u_0}{\dif v_0}=-\frac{u_0(u_0+v_0-3)}{v_0(u_0-1)} .
\label{equation:u0:v0}
\end{equation}

Furthermore, it can be shown \cite{Padmanabhan:1990,Chavanis:2002:a} that the inverse temperature and the potential energy can be written as
\begin{align}
\frac{1}{T}&=\frac{R}{|\kappa|N}v_0 ,\label{temperature:gravity}\\
W&=NT(u_0-3) .\label{potential:gravity}
\end{align}
What we need is to express the potential energy as a function of $T$, $P$, and $\mu$ only.
Since $\Phi(R)=\kappa N/R$, using (\ref{chemical:potential}) yields 
\begin{equation}
N=\frac{R}{|\kappa|}\left[T\ln\left(\frac{P\lambda_T^3}{T}\right)-\mu\right] , 
\label{N:gravity}
\end{equation}
where we have taken into account the local equation of state so that $P=n(R)T$.
From (\ref{N:gravity}) and (\ref{temperature:gravity}) one sees that
\begin{equation}
v_0(T,P,\mu)=\ln\left(\frac{P\lambda_T^3}{T}\right)-\frac{\mu}{T}
\label{v0}
\end{equation}
and therefore
\begin{equation}
\left(\frac{\partial v_0}{\partial T}\right)_{P,\mu}=\frac{1}{T}\left(\frac{\mu}{T}-\frac{5}{2}\right) .
\label{v0:T} 
\end{equation}
Combining (\ref{potential:gravity}) and the global equation of state (\ref{equation:state}) one obtains $3PV=NTu_0$ and hence, taking it into account and using (\ref{temperature:gravity}) to express $N$, the radius can be written as
\begin{equation}
R=T\left(\frac{v_0u_0}{4\pi|\kappa|P}\right)^{1/2} .
\label{R:gravity}
\end{equation}
Using (\ref{temperature:gravity}) and (\ref{R:gravity}), the potential energy (\ref{potential:gravity}) takes the form
\begin{equation}
W(T,P,\mu)=\frac{T^3 v_0^{3/2}}{\left(4\pi|\kappa|^3P\right)^{1/2}}\left(u_0^{3/2}-3u_0^{1/2}\right) , 
\end{equation}
which depends only on the desired variables since $u_0=u_0\left(v_0(T,P,\mu)\right)$.
We then have
\begin{equation}
\begin{split}
\left(\frac{\partial W}{\partial T}\right)_{P,\mu}=&\frac{3W}{T}\\
&+\frac{3W}{2v_0}\left(\frac{\partial v_0}{\partial T}\right)_{P,\mu}\left[1+\frac{v_0(u_0-1)}{u_0(u_0-3)}\frac{\dif u_0}{\dif v_0}\right] .
\end{split}
\label{derivative:potential}
\end{equation}
Therefore, according to (\ref{Gibbs-Duhem:relations}) and using (\ref{equation:u0:v0}), (\ref{potential:gravity}) and (\ref{v0:T}), from Eq.~(\ref{derivative:potential}) the MF entropy is obtained,
\begin{equation}
S= -N\left[\frac{\mu}{T}-2u_0+\frac{7}{2} \right] .
\end{equation}
Using that the thermal wavelength can be expressed as
$\lambda_T=\left[ 2\pi\hbar^2R v_0/(N|\kappa|m)\right]^{1/2}$
and that $P/T=Nu_0/(3V)$, from (\ref{v0}) one gets
\begin{equation}
\frac{\mu}{T}=\frac{1}{2}\ln(v_0)+\ln(v_0u_0)-v_0-\frac{1}{2}\ln\left(\frac{2NR^3m^3|\kappa|^3}{\pi\hbar^6}\right) , 
\end{equation}
and thus the entropy becomes
\begin{equation}
S= N\left[v_0 +2u_0-\frac{1}{2}\ln(v_0)-\ln(v_0u_0)- 3 \right]+S_0 ,
\end{equation}
as given in \cite{Chavanis:2005}, with
$S_0=N/2\ln\left[2NR^3m^3|\kappa|^3/\left(\pi\e\hbar^6\right)\right]$.
This verifies the first of the relations (\ref{Gibbs-Duhem:relations}) for the self-gravitating gas and the remaining two follow from an analogous procedure.

As stated in the previous sections, the validity of the MF depends on the ensemble considered. The density contrast can be written in terms of the solution of the Emden equation, yielding $\mathcal{R}=\exp\left(\psi_0\right)$. The stability in the microcanonical ensemble is restricted to $\mathcal{R}<709$, in the canonical ensemble these systems are stable for $\mathcal{R}<32.1$, and for $\mathcal{R}<1.58$ in the grand canonical ensemble.

\subsection{Small systems}

Systems with a small number of constituents can be treated from a thermodynamic point of view by considering them as independent members of an ensemble.
Since such an ensemble of independent small systems is a macroscopic system itself where the standard thermodynamics applies, thermodynamic properties of a single small system can be derived.
This is the well known approach introduced by Hill \cite{Hill:1963}.
The basic idea is to consider a chemical potential, denoted by $\mathscr{E}$, that accounts for the energy gained by the system when the number of members of the ensemble, $\mathscr{N}$, varies.
The energy, volume, number of particles and entropy of each small system are $E$, $V$, $N$, and $S$, respectively, while the total magnitudes corresponding to the whole ensemble are given by
$E_t=\mathscr{N}E$, $V_t=\mathscr{N}V$, $N_t=\mathscr{N}N$, and $S_t=\mathscr{N}S$.
Since the whole ensemble is a macroscopic system, these magnitudes satisfy
\begin{equation}
T\dif S_t=\dif E_t+P\dif V_t-\mu\dif N_t- \mathscr{E}\dif \mathscr{N}
\end{equation}
and hence,
$TS_t=E_t+P V_t-\mu N_t- \mathscr{E}\mathscr{N}$.
Thus, for any small system:
\begin{subequations}
\begin{align}
&TS=E+P V-\mu N- \mathscr{E} ,\\
&T\dif S=\dif E+P\dif V-\mu\dif N ,\\
&\dif \mathscr{E}=-S\dif T+V\dif P-N\dif \mu .
\end{align}
\label{Hill:relations}
\end{subequations}
Hill mentioned that the use of different environmental variables, i.e., control parameters, would lead to different descriptions of the thermodynamic phenomena when small systems are considered \cite{Hill:1963}.
In this case, ensemble inequivalence is due to the finite size of the systems.
In fact, negative heat capacity leading to ensemble inequivalence is a generic feature of finite systems at phase coexistence \cite{Chomaz:2002}.
More specifically, a first-order phase transition in small systems is associated with a convexity anomaly in the appropriate thermodynamic potential \cite{Chomaz:2002}. This convexity anomaly is a signal of ensemble inequivalence.
In addition, systems with long-range interactions can be regarded as finite systems as well, in the sense that the range of the interaction is comparable with the size of the system \cite{Chomaz:2002}.

Our aim here is to point out another common feature between small systems and systems with long-range interactions.
From the beginning we have been considering thermodynamic properties of a single system.
Even so, if ones relates the energy $\mathscr{E}$ to the potential energy such that $\mathscr{E}= -\sigma W$, Eqs.~(\ref{Hill:relations}) become the same equations we have found for systems with long-range interactions.
Although $W$ and $\mathscr{E}$ are introduced for different reasons, they formally play the same role.
Therefore, a small system such as a macromolecule and a system with long-range interactions such as a star, both being finite system, can be formally described by the same thermodynamic relations.

\section{Discussion}
\label{Discussion}

We have studied the local thermodynamics of systems with strong long-range interactions in $d$ dimensions by computing local thermodynamic potentials per particle.
We have considered pair interaction potentials that decay as $1/r^\nu$, with $0\leq\nu\leq d$.
By computing the local entropy per particle and using the condition of local thermodynamic equilibrium we have obtained the local equation of state, which corresponds to the isothermal ideal gas equation with the density depending on the long-range interaction potential that couples the constituents of the system.
This result coincides with the results that were obtained in previous works where the condition of hydrostatic equilibrium was assumed \cite{deVega:2002:b,Chavanis:2011}.
Thus we have shown that interactions locally play the role of an external field that perturbs the gas and that the local entropy per particle is a function of the local kinetic energy and local volume: It is a Sackur-Tetrode-like entropy that implicitly depends on the interaction potential trough the local volume.

Different ensemble representations possess, in general, different ranges of validity in the space of parameters specifying the state of the system.
Since any global thermodynamic potential computed in different ensemble representations has the same functional form in the MF limit, local thermodynamic potentials also have the same functional form in the different ensemble representations.
We stress that the MF description breaks down at certain critical points that are specific to each ensemble. Beyond one of these critical points, the system can no longer be described with the MF in the corresponding ensemble, but still may be well described in a different ensemble.

By volume integration of the relation satisfied by the thermodynamic magnitudes at the local level, we have obtained the equation satisfied by the set of global magnitudes.
Remarkably, the potential energy enters as a thermodynamic variable that modifies the global thermodynamic potentials.
In the thermodynamics of systems with short-range interactions the potential energy is obviously involved since it is included in the total energy, but it does not appear as an additional contribution that modifies the thermodynamic potentials.
Such a contribution is proportional to $\sigma=(d-\nu)/\nu$ and hence, for the marginal case $\nu=d$, the usual thermodynamic relations valid for short-range potentials are recovered.

As a consequence of this contribution coming from the total potential energy, the Gibbs-Duhem equation is modified and the variables $T$, $P$, and $\mu$ become independent if long-range interactions are present in the system.
By taking advantage of the thermodynamic relations obtained from this generalized Gibbs-Duhem equation, the entropies of a system with a spatially uniform potential and of the self-gravitating gas have been computed.

It is important to note that the Euler relation (\ref{global:relation:2}) reveals that the energy $E$ is not a linear homogeneous function of $S$, $V$, and $N$ if interactions are long ranged. This was already stated by Hill in his study of small systems \cite{Hill:1963}, which, as noted previously, satisfy the same kind of thermodynamic equations.
Long-range interactions introduce an extra degree of freedom that causes this deviation from the usual scaling law of systems with short-range interactions.
This extra degree of freedom appears in (\ref{global:relation:2}) as the term $\sigma W$.

To derive the relations between thermodynamic quantities at the local level it is not necessary to make the form of the interaction potential explicit; it suffices to assume that the potential is long ranged.
At the global level, however, the interaction potential determines the equation of state and hence all global thermodynamic relations.
Since we considered explicitly the virial relation for a power-law interaction, the obtained results are valid for this case only and other cases require individual attention.
In the framework of thermodynamics of systems with short-range interactions, the Gibbs-Duhem equation plays a central role in the sense that it provides a valuable tool to derive useful relations that are implemented in the solution of thermodynamic problems.
In the general case of arbitrary long-range interactions, the usual Gibbs-Duhem equation cannot be taken for granted, but the analogous equation must be derived.

Succinctly, the Gibbs-Duhem equation and the Euler relation we obtained are valid for systems with kinetic degrees of freedom and interacting potentials of the form $1/r^\nu$, $0\leq\nu\leq d$.
Therefore, the cases of gravity and Coulomb interactions in two dimensions \cite{Teles:2010,Levin:2008:b} are not included.
Trapped non-neutral plasmas are an important example of the latter \cite{Levin:2008:b}, which provide the possibility of testing long-range interacting systems in a laboratory.
Furthermore, confinement in this case is accomplished by using an external field instead of rigid walls, so another difference with the treatment given here can be noted: We considered ensembles that all have the volume as a control parameter.
If fluctuations in $V$ are allowed, the corresponding ensemble with adequate control parameters must be utilized.

\begin{acknowledgments}
We thank the anonymous referees for helpful comments.
One of the authors (I.L.) thanks Markus Fr\"ob for fruitful discussions. He also acknowledges financial support through an FPI scholarship (Grant No. BES-2012-054782) from the Spanish Government.
This work was supported by the Spanish Government under Grant No. FIS2011-22603.
\end{acknowledgments}

\appendix*

\section{}
\label{appendix}
In this appendix we obtain (\ref{integral}) as a functional integral over the number density.
We employ a general method to derive the MF description that was previously used in the context of long-range interacting systems to treat Newtonian gravity in $d=3$ \cite{Thirring:1970,deVega:2002:a}.
To implement this method, the $d$-dimensional volume $V$ is divided in $M$ cells, $1\ll M\ll N$, of volume $v_a=V/M$, so that
$\sum_{a=1}^M v_a=V$.
Each cell is located at
$\vect{x}_a=(x_a^1,x_a^2,\dots,x_a^d)$
and contains $n_a\gg 1$ particles such that
$\sum_{a=1}^M n_a=N$.
This construction is self-consistent if the number of particles is large enough, i.e., the large $N$ limit is implicitly assumed.
It is also assumed that the size of each cell is small enough so that the potential can be considered constant through the cell, but, of course, it varies from cell to cell.
It follows that the total potential energy can be approximated as
$W\approx\sum_{a> b}^M n_a \phi(\vect{x}_a,\vect{x}_b) n_b$.

Furthermore, with the help of the multinomial theorem it is not difficult to derive the following identity:
\begin{equation}
\frac{1}{N!} \int  \frac{\dif^{dN}\vect{q}}{V^N}=\sum_{\{n_1,\dots, n_M\}}\delta_{N,\sum\limits_{i=1}^M n_i} \prod_{i=1}^M
\frac{1}{n_i!}\left(\frac{v_i}{V}\right)^{n_i} ,
\label{summation}
\end{equation}
where $\{n_1,\dots, n_M\}$ means all possible values $n_1,n_2,\dots,n_M$ of the occupation numbers and the Kronecker $\delta$ restricts the total number of particles to $N$.
Equation~(\ref{summation}) can be seen as a way to perform an integral in the $dN$-dimensional configuration space by summing over all possible occupation number distributions through the cells of a single-particle configuration space.
Thus, using the multinomial expansion, one approximates (\ref{integral}) by
\begin{widetext}
\begin{equation}
\e^{J(\beta)}\approx \sum_{\{n_1,\dots, n_M\}}\delta_{N,\sum\limits_{i=1}^M n_i} \prod_{i=1}^M\frac{1}{n_i!} \left(\frac{v_i}{V}\right)^{n_i}\exp\left(-\beta \sum_{a>b}^M n_a n_b \phi(\vect{x}_a,\vect{x}_b)-\frac{d}{2}\ln\left(\varepsilon\beta\right)\sum\limits_{a=1}^M n_a\right) .
\label{exp:J:appendix}
\end{equation}
\end{widetext}
The above approximation formulated in a discrete way allows one to obtain a continuous field representation.
This continuum limit is obtained by introducing the number density, whose value in each cell is given by $n(\vect{x})=n_a/v_a$.
Consequently, summations over the cells become spatial integrals and the summation over occupation number distributions becomes a functional integration in the number density:
\begin{equation*}
\sum_{a=1}^M \rightarrow\int \frac{\dif^d\vect{x}}{v_a}\qquad \mbox{and} \qquad\sum_{\{n_1,\dots, n_M\}}\rightarrow \int\mathcal{D}n(\vect{x}) .
\end{equation*}
With Stirling's approximation we also have
\begin{equation*}
\prod_{i=1}^M \frac{1}{n_i!}\left(\frac{v_i}{V}\right)^{n_i}= \exp\left(-\int n(\vect{x})\ln \frac{n(\vect{x})V}{\e}\,
\dif^d\vect{x}\right) .
\end{equation*}
In addition, the Kronecker $\delta$ restricting the number of particles becomes a Dirac $\delta$ in the continuum limit and thus can be represented in the form
$\delta\left(x\right)=(2\pi\iunit)^{-1} \int_{-\iunit\infty}^{\iunit\infty}\dif\alpha\, \e^{-\alpha x}$. Therefore, taking these prescriptions into account, Eq.(\ref{exp:J:appendix}) can be rewritten as Eq. (\ref{exp:J}) in the MF approximation.

\bibliography{mainbib}

\end{document}